\documentclass[twocolumn,fp]{jpsj3}
\usepackage{txfonts}

\title{Ferromagnetic Cluster-Glass State in Itinerant Electron System Sr$_{1-x}$La$_x$RuO$_3$}

\author{Ikuto~Kawasaki$^{1}$\thanks{E-mail address: kawasaki{\_}ikuto@riken.jp}, Makoto~Yokoyama$^{2}$\thanks{E-mail address: makotti@mx.ibaraki.ac.jp}, Suguru~Nakano$^{2}$,  Kenji~Fujimura$^{2}$, Naoaki~Netsu$^{2}$, Hirofumi~Kawanaka$^{3}$,  and Kenichi~Tenya$^{4}$  
}

\inst{\address{$^{1}$Advanced Meson Science Laboratory, RIKEN Nishina Center for Accelerator Based Science, RIKEN, Wako 351-0198, Japan} \\
\address{$^{2}$Faculty of Science, Ibaraki University, Mito 310-8512, Japan} \\
\address{$^{3}$National Institute of Advanced Industrial Science and Technology, Tsukuba 305-8568, Japan} \\
\address{$^{4}$Faculty of Education, Shinshu University, Nagano 390-8621, Japan} \\

}

\abst{The magnetic and electronic properties of  Sr$_{1-x}$La$_x$RuO$_3$ were studied by means of dc-magnetization, ac-susceptibility, specific heat, and electrical resistivity measurements. The dc-magnetization and ac-susceptibility measurements have revealed that the transition temperature and the ordered moment of the ferromagnetic order are strongly suppressed as La is substituted for Sr. The ac-susceptibility exhibits a peak at $T^{*}$ due to the occurrence of spontaneous spin polarization. Furthermore, we observed that $T^{*}$ shows clear frequency variations for $x$ $\geq$ 0.3. The magnitude of the frequency shifts of  $T^{*}$ is comparable to that of cluster-glass systems, and the frequency dependence is well described in terms of the Vogel-Fulcher law. On the other hand, it is found that the linear specific heat coefficient $\gamma$ enhances with the suppression of the ferromagnetic order. The relatively large $\gamma$ values reflect the presence of the Ru 4$d$ state at Fermi level, and hence, the magnetism of this system is considered to be tightly coupled with the itinerant characteristics of the Ru 4$d$ electrons. The present experimental results and analyses suggest that the intrinsic coexistence of the spatially inhomogeneous magnetic  state and the itinerant nature of the Ru 4$d$ electrons is realized in this system,  and such a feature may be commonly involved in La- and Ca-doped SrRuO$_3$. }

\begin{document}
\maketitle

\section{Introduction}

Ruthenium-based oxides exhibit various types of interesting properties, such as  quantum criticality,\cite{gri,gri2}  non-Fermi liquid behavior,\cite{khali} and unconventional superconductivity.\cite{maeno} These properties originate from the Ru 4$d$ orbitals, which are more extended than the 3$d$ orbitals in 3$d$ transition metals and  are  therefore expected to have  an itinerant character. It is considered from observations of the above intriguing properties that strong electronic correlation between the Ru 4$d$ electrons plays crucial roles in  Ruthenium-based oxides.

SrRuO$_3$ crystallizes into a distorted perovskite structure and is a ferromagnet with a Curie temperature of about 160 K, whose ordered moment is about 1.1 $\mu_\mathrm{B}$.\cite{calla,kanba}  Photoemission experiments showed that the density of states at Fermi level is dominated by the Ru 4$d$ state and the overall distribution of the Ru $4d$ and O 2$p$ states are well reproduced by band structure calculations.\cite{fujio_ph,okamoto_ph,park_ph,kim_ph,taki_ph,siemon_ph,grebin_ph} Itinerant Ru 4$d$ states are thus considered to be responsible for the magnetic properties. It is also argued that the development of the incoherent component in the density of states reflects the electronic correlation effects.
In addition, optical studies revealed that the charge dynamics of this system are significantly different from those expected in usual Fermi liquid systems.\cite{kostic,doge2} Another remarkable feature of this compound is ``bad metal" behavior in transport at high temperatures: the electrical resistivity continues to increase with increasing temperature, even though the Boltzmann mean free path becomes smaller than the lattice constants, indicating that the itinerant quasi-particle description is no longer available in the high temperature range.\cite{allen,gunn} These experimental findings suggest that the magnetic and the transport properties are strongly influenced by the correlation of the Ru 4$d$ electrons, and the Ru 4$d$ states have a duality of itinerant and localized natures.

In fact, the itinerant ferromagnetic (FM) state in SrRuO$_3$ has an instability toward electron localization because of the effect of a strong electronic correlation. Such electron localization can be found in the Ru site-substituted system SrRu$_{1-x}$Mn$_x$O$_3$, where doping Mn suppresses  the itinerant FM state, and an insulating phase coexisting with a localized antiferromagnetic order appears above the critical concentration $x_c$ = 0.39.\cite{cao,sahu,yoko,kole} Similar insulating phases have been reported for several Ru site-substituted systems.\cite{kwkim,williams}

Suppression of the ferromagnetism has also been found in the Sr site-substituted system Sr$_{1-x}$Ca$_x$RuO$_3$.\cite{gcao,kiyama,yoshimura,okamoto,demko,hosaka} Substituting Ca for Sr  increases the Ru-O-Ru bond angle and thereby enhances the RuO$_6$ octahedra rotation in the distorted perovskite structure. This variation is expected to strongly affect the electronic structure.\cite{mazin} Photoemission and x-ray absorption studies showed that  Ca substitution enhances the electronic correlation, and induces intensity transfer from the coherent to incoherent parts in the photoemission spectrum.\cite{park_ph,taki_ph} However, in contrast to SrRu$_{1-x}$Mn$_x$O$_3$, this system does not show a transition to an insulating phase but keeps a metallic character up to $x$ = 1. A recent magnetization study has revealed that the ferromagnetic-to-paramagnetic quantum phase transition driven by the substitution is totally destroyed by the disorder, and the FM phase is extended to over a wide $x$ range.\cite{demko} It is thus considered that both the electronic correlation and disorder play an important role in this system.

Sr$_{1-x}$La$_x$RuO$_3$ also shows a suppression of ferromagnetism generated by substituting La for Sr.\cite{bouchard,nakatsu} The La substitution also enhances the RuO$_6$ octahedra rotation.\cite{nakatsu} In addition, the Ru-O distance increases with increasing $x$, suggesting that doping La may enhance the role of the electronic interaction through the  suppression of the Ru-O hybridization.\cite{nakatsu} Therefore, we expected that both the electronic correlation and disorder play important roles in this compounds, and it would be a subject of interest.  In order to shed light on this point, we have studied the magnetic and electronic properties of Sr$_{1-x}$La$_x$RuO$_3$ by means of dc-magnetization, ac-susceptibility, electrical  resistivity, and specific heat measurements.  Recently, we reported preliminary investigations of the magnetic and electronic properties.\cite{sces}  In the present paper, we will show the detailed experimental results and discussion about the magnetic properties and the nature of the Ru 4$d$ electrons.

\section{Experimental Details}

The polycrystalline samples of Sr$_{1-x}$La$_x$RuO$_3$ for $x$ $\leq$ 0.5 were synthesized by the conventional solid-state reaction method with starting materials of SrCO$_3$, La(OH)$_3$, RuO$_2$, and Ru. The powders of these materials with stoichiometric compositions were mixed and  calcined at 1000 ${}^\circ\mathrm{C}$ for 6 hours. After careful mixing of the calcined samples, they were shaped into pellets and then sintered at 1000 ${}^\circ\mathrm{C}$ for 24-48 hours. X-ray diffraction measurements confirmed that all samples crystallize in a distorted perovskite structure (the GaFeO$_3$-type orthorhombic structure), without any extrinsic phase within the experimental accuracy. However, ac-susceptibility measurements for the La-doped sample revealed a slight contamination of a FM impurity phase of pure SrRuO$_3$. We found that such an impurity component can be reduced by iterating the above sinter process several times; for the present experiment we used  samples whose impurity components were well reduced down to a few percent in volume.   
The specific heat measurements were performed  in the temperature range of 3.2~K $<$ $T$ $<$ 275~K by means of a thermal-relaxation method.  The ac-susceptibility was measured in the temperature range of 4~K $<$ $T$ $<$ 300~K with an applied ac-field of $\sim$1 Oe using a standard Hartshorn bridge circuit. The frequency of applied ac-field ranges from 7 Hz to 1520 Hz. The dc-magnetization measurements were carried out using a commercial SQUID magnetometer in the temperature range of 5-300 K. The electrical resistivity measurements were performed between 3 K and 300 K by the conventional four-probe method.

\section{Results}

\begin{figure}[tbp]
\begin{center}
\includegraphics[keepaspectratio, width=7cm,bb = 0 15 280 700,clip]{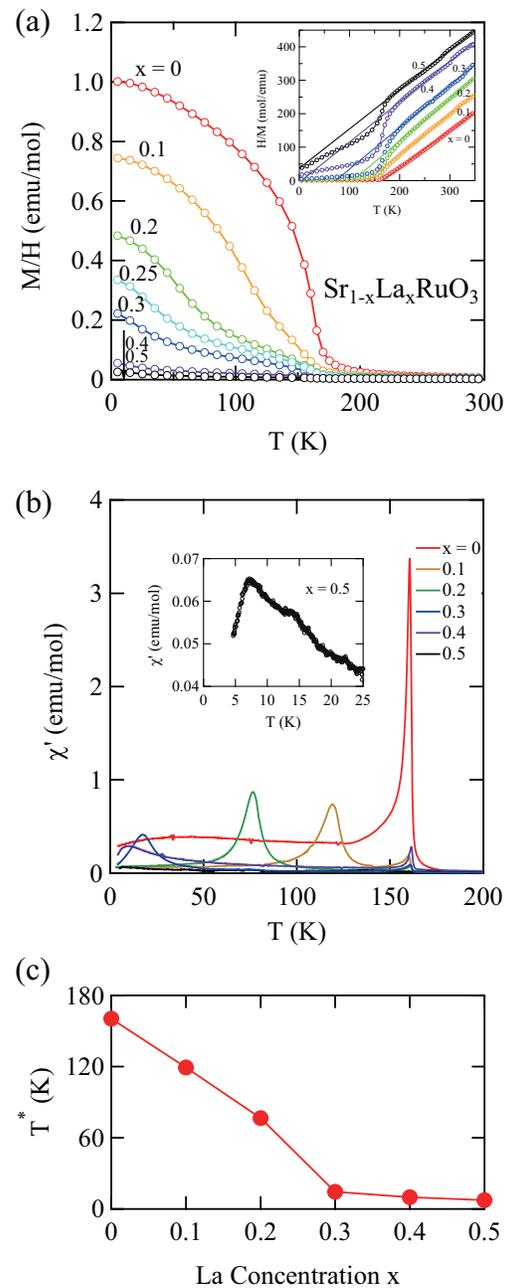}
\end{center}
\caption{(Color online) (a) Temperature dependence of the dc-magnetization for Sr$_{1-x}$La$_x$RuO$_3$. The inset shows the temperature dependence of the inverse susceptibility. The solid lines represent the best fits using the Curie-Weiss law. (b)  Temperature dependence of the in-phase component of the ac-susceptibility measured at a frequency of 180 Hz. The inset displays the enlargement for $x$ = 0.5. (c) Onset of the FM order $T^{*}$ estimated from the peak in  $\chi'_\mathrm{ac}$, plotted as a function of the La concentration $x$. 
}
\label{f1}
\end{figure}

Figure 1(a) shows the temperature dependence of the dc-magnetization of Sr$_{1-x}$La$_x$RuO$_3$ for $x$ $\leq$ 0.5 measured under a field of 5 kOe and field-cooled condition.   The transition temperature and  the ordered moment of the FM order are strongly suppressed with increasing La concentration $x$. The sharp increase in the magnetization at the FM ordering becomes broad by doping La, which is presumably due to the effect of disorder generated by doping La. The inverse susceptibility is plotted as a function of temperature in the inset of Fig. 1(a). At high temperatures, the susceptibility obeys the Curie-Weiss law given by $\chi^{-1}$ = $(T-\Theta)/C$, where $C$ and $\Theta$ are the Curie constant and the Curie-Weiss temperature, respectively. The solid lines in the inset of Fig. 1(a) represent the best fits using the Curie-Weiss law in the fitting range of $T$ $\geq$ 200 K. $\Theta$ decreases with $x$, and then becomes slightly negative for $x$ = 0.5, indicating a suppression of the FM interaction. These results are consistent with previous reports.\cite{bouchard,nakatsu}

We performed ac-susceptibility measurements  in order to investigate the $x$ dependence of the ordering temperature. The in-phase components of the ac-susceptibility $\chi'_\mathrm{ac}$ at a frequency of 180 Hz for $x$ $\leq$ 0.5 are shown in Fig. 1(b). $\chi'_\mathrm{ac}$ of SrRuO$_3$ exhibits a clear and sharp peak at the FM transition temperature of 161 K. We here define $T^{*}$ as the temperature of the peak seen in the temperature variations of $\chi'_\mathrm{ac}$.  The $T^{*}$ values and the magnitude of $\chi'_\mathrm{ac}$ at $T^{*}$ are significantly suppressed with increasing $x$. A development of a peak is also observed in the out-phase component of ac-susceptibility $\chi''_\mathrm{ac}$. For each La concentration, it is found that the peak-top temperature is slightly lower than $T^{*}$, and the magnitude of the peak is approximately two order smaller than that seen in $\chi'_\mathrm{ac}$. However,  we could not perform detailed analyses on the peak in $\chi''_\mathrm{ac}$ because of the very large difference of the peak heights between $\chi'_\mathrm{ac}$ and $\chi''_\mathrm{ac}$ as well as the smallness of the $\chi''_\mathrm{ac}$ peak itself.

We further observed that $\chi'_\mathrm{ac}$ for $x$ $\geq$ 0.1 shows a small peak at 161 K. This probably originates from the presence of a fragmentary phase of the pure SrRuO$_3$. The sharpness of the peak due to the fragmentary phase is always comparable to that for pure SrRuO$_3$, though their magnitude is highly reduced and depends on the sample preparation process. In particular, we found that the magnitude is effectively reduced by iterating the sinter process, while the characteristics of the peak at $T^{*}$ hardly depends on the sample preparation process. This reflects that the fragmentary phase has a bulk property.
The FM ordering in this phase occurs at much higher temperature than $T^{*}$ in the La substitutions, and hence, it is expected that the fragmental FM order becomes stable, and the dynamic characteristics of the FM spins in this phase are strongly reduced below  $\sim T^{*}$.  In addition, its volume fractions, estimated from the peak heights, are only a few percent, and become small as the La concentration is increased. Thus, it is not likely that such a small fragmentary component significantly affects the magnetic properties of the majority phase at low temperatures. Figure 1(c) shows the $x$ dependence of $T^{*}$.  $T^{*}$ monotonously decreases with increasing $x$ up to $x$ = 0.3, and becomes very small but never reaches zero for $x$ $\leq$ 0.5.   

\begin{figure}[tbp] 
\begin{center}
\includegraphics[keepaspectratio, width=7.5cm,bb = 0 0 300 490,clip]{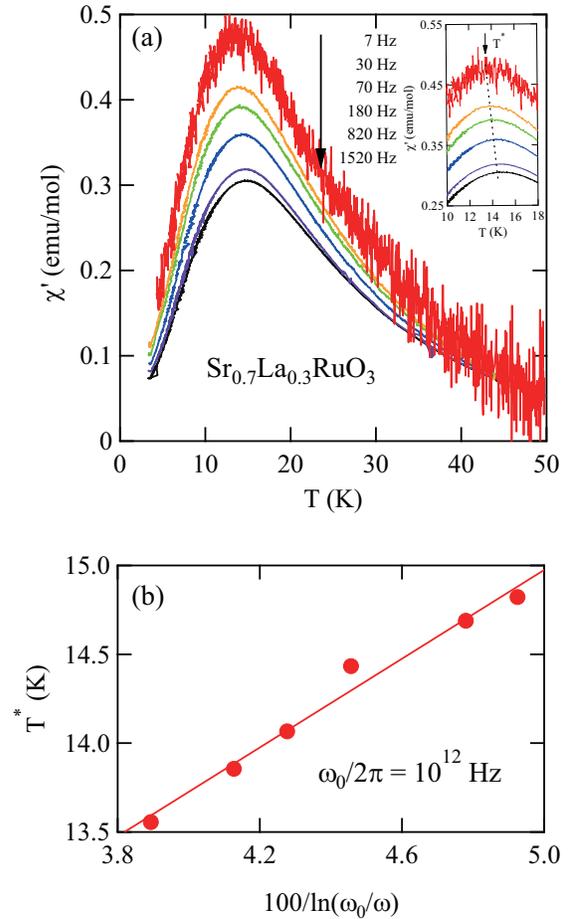}
\end{center}
\caption{(Color online) (a) In-phase components of the ac-susceptibility measured under ac-fields with various frequencies for $x$ = 0.3.   The inset shows an enlargement around the peak temperatures $T^{*}$. (b) Frequency variation of  $T^{*}$ plotted as  $T^{*}$ versus 100/ln($\omega_0/\omega$). The solid line is a fit using the Vogel-Fulcher law. }
\label{f1}
\end{figure}

\begin{figure}[tbp]
\begin{center}
\includegraphics[keepaspectratio, width=7.3cm,bb = 0 0 270 410,clip]{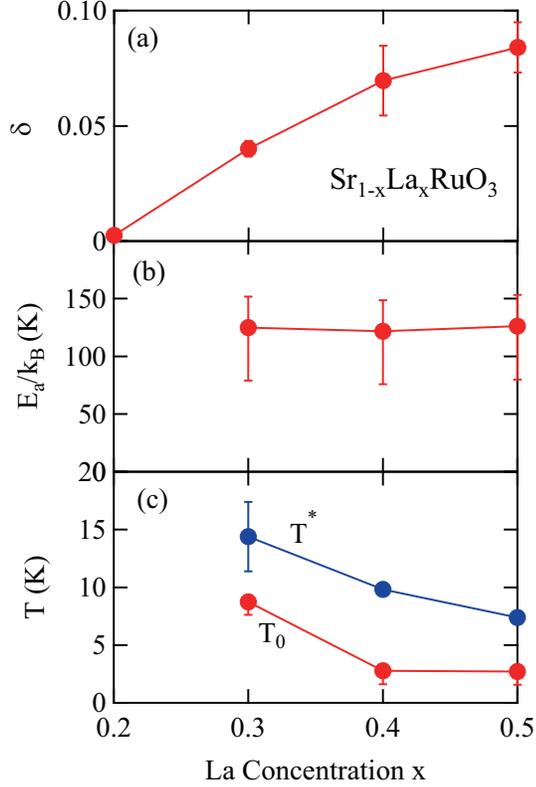}
\end{center}
\caption{(Color online) La concentration dependence of  (a) the initial frequency shift $\delta$, (b) the activation energy $E_\mathrm{a}$, (c) the Vogel-Fulcher temperature $T_\mathrm{0}$, and the freezing temperature $T^{*}$. $E_\mathrm{a}$ and $T_\mathrm{0}$ are derived by assuming $\omega_0/2\pi$ to be $10^{12}$ Hz. The error bars for  $E_\mathrm{a}$ and $T_\mathrm{0}$ correspond to the $\omega_0/2\pi$ range from $10^{10}$ to $10^{13}$ Hz.}
\label{f1}
\end{figure}

$\chi'_\mathrm{ac}$ for $x$ $\leq$ 0.2 still shows a relatively sharp peak at  $T^{*}$. On the other hand, the peak width of $\chi'_\mathrm{ac}$  becomes significantly broad for $x$ $\geq$ 0.3, accompanying a reduction of the intensity  of $\chi'_\mathrm{ac}$ at $T^{*}$. This implies that the occurrence of the peak in $\chi'_\mathrm{ac}$ involves a glassy magnetic nature. We investigated the  frequency dependence of the ac-susceptibility  in order to confirm this. The in-phase components of the ac-susceptibility for  $x$ = 0.3 measured at various frequencies are plotted in Fig. 2(a). With increasing frequency $\omega$,   $T^{*}$ moves to higher temperatures in connection with a decrease of the magnitude  of $\chi'_\mathrm{ac}$ at $T^{*}$. We observed a similar frequency dependence of  $T^{*}$ for the $\chi'_\mathrm{ac}$ data of $x$ $\geq$ 0.2. It is well known that  $T^{*}$ does not shift with $\omega$ in such a low frequency range when a normal FM order occurs, and the shift can usually be observed in the frequency range of MHz to GHz.\cite{mydosh}  
 We here estimate the initial frequency shift $\delta$ =  $\Delta T^{*}/(T^{*}\Delta \mathrm{log_{10}}\omega$), by which one can compare the frequency sensitivity of  $T^{*}$ in different systems. The $\delta$ values evaluated  for Sr$_{1-x}$La$_x$RuO$_3$ are shown in Fig. 3(a), and relatively large $\delta$ values are obtained for $x$ $\geq$ 0.3. The $\delta$ values range from 0.040 ($x$ = 0.3) to 0.084 ($x$ = 0.5), which are larger than those reported for the canonical spin-glass systems, e.g., $\delta\sim$ 0.005 (CuMn) whereas are smaller than those reported for non-interacting ideal superparamagnetic systems ($\delta\sim$ 0.1).\cite{mydosh, dormann} Instead, the $\delta$ values for 0.3 $\leq x\leq$ 0.5 are comparable to those of cluster-glass systems, which are realized as an ensemble of interacting magnetic clusters.

Since the cluster-glass behavior is suggested for $x\geq$ 0.3, thermal activation processes of magnetic clusters are considered to be closely related to the frequency dependence of $\chi'_\mathrm{ac}$. In order to characterize the freezing process of the magnetic clusters, we analyze the  $T^{*}$ data using the Vogel-Fulcher law\cite{mydosh,vogel,fulcher} given by 
\begin{eqnarray}
\omega=\omega_0\mathrm{exp}\biggl(-\frac{E_\mathrm{a}}{k_\mathrm{B}(T^{*}-T_0)}\biggr), 
\end{eqnarray}   
where $\omega_0$ is the attempt frequency of the clusters, $E_\mathrm{a}$ is the activation energy, and 
$T_0$ is the Vogel-Fulcher temperature, which is often considered to be a measure of the  strength of the intercluster interactions. For analyzing the data, it is useful to rewrite  Eq. (1) as
\begin{eqnarray}
T^{*}=T_0+\frac{E_\mathrm{a}}{k_\mathrm{B}}[\mathrm{ln}(\frac{\omega_0}{\omega})]^{-1}.
\end{eqnarray}
$\omega_0/2\pi$ is expected to have values ranging from $10^{10}$ to $10^{13}$ Hz for the typical magnetic cluster systems, and we here fixed it to be 10$^{12}$ Hz for the present analysis.\cite{mydosh} The best fit for $x$ = 0.3 is shown in Fig. 2(b). The $T^{*}$ data well obey the Vogel-Fulcher law.  The $x$ dependence of  $E_\mathrm{a}$ and $T_0$ obtained by the above analysis are plotted in Figs. 3(b) and 3(c), where the vertical error bars correspond to the $\omega_0/2\pi$ range from $10^{10}$ to $10^{13}$ Hz. For a comparison, $T^{*}$ is also shown in Fig. 3(c). The $E_\mathrm{a}$ values,  which represent the energy barrier of the cluster flipping, hardly depend on $x$. In contrast, the reduction of $T_0$ for 0.3 $\leq$ $x$ $\leq$ 0.4 can be ascribed to a suppression of the intercluster interaction. We thus consider that the decrease in $T^{*}$ is mainly due to the suppression of the intercluster interaction  for $x$ $\geq$ 0.3. The small $T_0$ values observed for $x$ = 0.4 and 0.5 seem to be connected with the fact  that the frequency variation of $T^{*}$ roughly obeys the Arrehenius law predicted for the non-interacting magnetic cluster systems, since the Vogel-Fulcher law reduces to the Arrehennius law  if  $T_0$ becomes zero.  This is consistent with the  large $\delta$ values observed for $x$ = 0.4 and 0.5, which are relatively close to those of superparamagnetic systems. Here, one might suspect that the fragmentary FM grains recognized by the $\chi'_\mathrm{ac}$ peak at 161 K affect the freezing process of the cluster-glass phases. However, we emphasize that $T^{*}$ of the fragmentary FM grains does not show any measurable frequency dependence in the ac-susceptibility. This indicates that the dynamical properties of these FM grains differ greatly from those seen in the majority cluster-glass phases. Furthermore, the FM impurity grains are considered to be static at $\sim$$T^{*}$, since their ordering temperature ($\sim$161 K) is much higher than  $T^{*}$. These properties of the FM impurity grains are clearly incompatible with the gradual freezing expected in the dynamical magnetic clusters. We thus consider that these FM grains cannot be an origin of the cluster-glass phases, and there is no coupling between them. 
 
\begin{figure}[tbp]
\begin{center}
\includegraphics[keepaspectratio, width=7.3cm,bb = 0 0 390 690,clip]{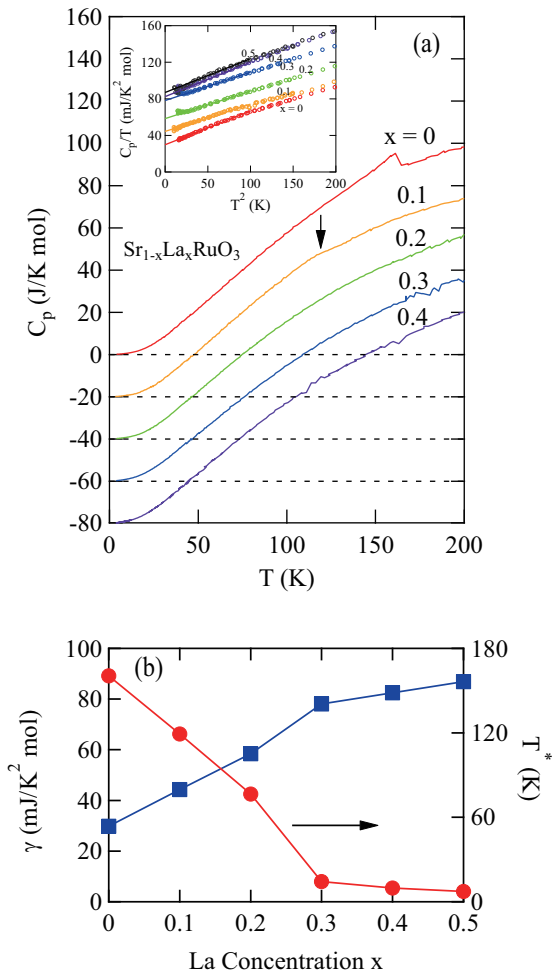}
\end{center}
\caption{(Color online) (a) Temperature dependence of the specific heat for Sr$_{1-x}$La$_x$RuO$_3$. The baselines of the data for $x$ $\geq$ 0.1 are transformed for clarity. The arrow indicates the position of $T^{*}$ for $x$ = 0.1. The specific heat data at low temperatures plotted as $C_p/T$ vs $T^2$ are shown in the inset, and the solid lines in the inset are fits using a $\gamma$ + $\beta T^2$ function in the fitting range of $T$ $\leq$ 10 K.  (b) La concentration dependence of the linear specific heat coefficient. The $T^{*}$ data are also shown  for a comparison.   }
\label{f1}
\end{figure}

In Fig. 4(a), we present the temperature dependence of the specific heat $C_p$ of Sr$_{1-x}$La$_x$RuO$_3$. $C_p$ of SrRuO$_3$ shows a clear jump  attributed to the FM transition at 161 K. The anomaly at $T^{*}$ is strongly suppressed with doping La. A shoulder-like anomaly appears at around $T^{*}$ for $x$ = 0.1, which is indicated by an arrow in Fig. 4(a). On the other hand,  the anomaly at  $T^{*}$ becomes too small and broad to be observed for $x$ $\geq$ 0.2. This is probably caused by the disorder effect and the small magnetic entropy associated with FM ordering, which is much less than that expected for the localized Ru spins.\cite{allen}

The inset of Fig. 4(a) displays the specific heat data at low temperatures plotted as $C_p/T$ versus $T^2$. We found that $C_p/T$ of Sr$_{1-x}$La$_x$RuO$_3$ shows a quadratic temperature dependence at low temperatures for all the $x$ range presently investigated. This suggests that the low-temperature specific heat is mainly ascribed to excitations of the Fermi-liquid quasiparticles and phonon contributions. However, it should be remembered that the development of the spin fluctuation and the FM clusters may also contribute to $C_p$ at $\sim T^{*}$ in the intermediate La concentrations, though there is no clear indication of the spin-wave contribution (such as a $T^{3/2}$ function) being dominant in $C_p$ at low temperatures. In this context, these spin-entropy contributions are expected to be suppressed and spread in a wide temperature range  in the intermediate La concentrations, because present $C_p/T$ shows no anomalous behavior around $T^{*}$ such as a non-Fermi-liquid divergence or a peak structure. Thus, we simply fitted the $C_p/T$ data below 10 K with a $\gamma$ + $\beta T^2$ function.  The obtained $\gamma$ values are shown in Fig. 4(b). For a comparison, the $x$ dependence of $T^{*}$ is also shown in this figure. The $\gamma$ value for SrRuO$_3$ is estimated to be 30 mJ/K$^2$  mol, which is roughly in agreement with previous studies.\cite{allen,gcao} It increases with increasing  $x$, and shows a tendency of  saturation to the value of about 80 mJ/K$^2$ mol for $x$ $\geq$ 0.3. 
 The large $\gamma$ values observed in La-doped SrRuO$_3$ reflect the presence of the Ru 4$d$ density of states at Fermi level, indicative of an itinerant character being involved in the Ru 4$d$ electrons and a presence of the electronic correlation effect.  We found that the increase in $\gamma$ coincides with the decrease in $T^{*}$ (Fig. 4(b)). It is considered that the increase of $\gamma$ is caused by the following possible variations of the electronic and magnetic state generated by doping La: (i) a suppression of the splitting of up and down spin bands, (ii) a development of the electronic correlation effect, and (iii) an enhancement of the magnetic entropy contribution due to the fluctuating spins and the FM clustering. Here, we point out that the saturated value of $\gamma$ is very close to that of the isostructural non-magnetic metal CaRuO$_3$ (73-82 mJ/K$^2$),\cite{gcao,kiyamaT,kikugawaN} which has a similar band structure with SrRuO$_3$.\cite{santig} This similarity implies that the La ions hardly change the density of state around Fermi level and the presence of the mechanism (i).

\begin{figure}[tbp]
\begin{center}
\includegraphics[keepaspectratio, width=7.3cm,bb = 0 0 270 530,clip]{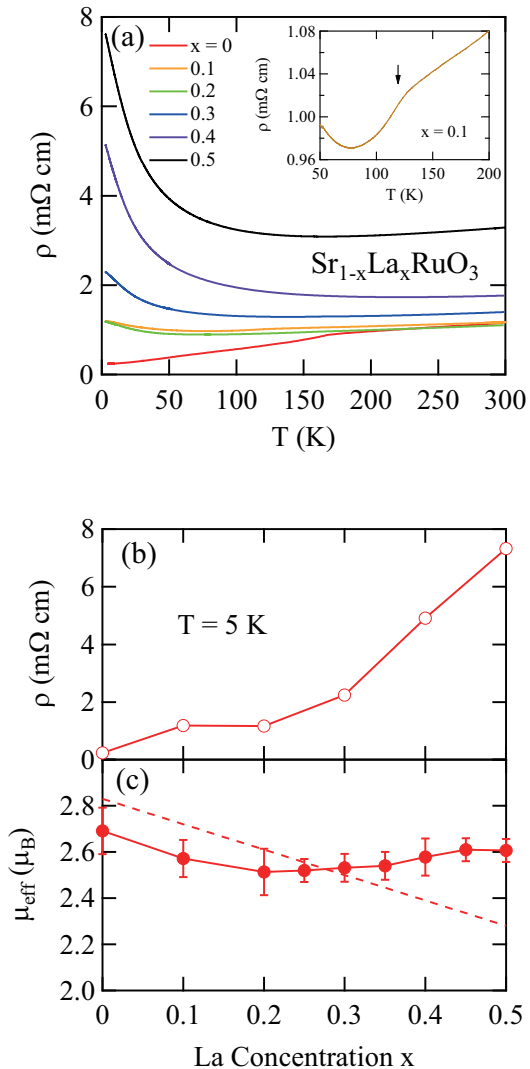}
\end{center}
\caption{(Color online) (a) Temperature dependence of the electrical resistivity for Sr$_{1-x}$La$_x$RuO$_3$. The inset displays the data for $x$ = 0.1, and the arrow  represents the position of $T^{*}$. The magnitude of the electrical resistivity at 5 K and the effective moment estimated from the Curie-Weiss fitting are also displayed in (b) and (c), respectively. The broken line in (c) represents the  effective moments calculated on the basis of a simple admixture of Ru$^{4+}$ and Ru$^{3+}$ ions.}
\label{f1}
\end{figure}

Figure 5(a) shows the temperature dependence of the electrical resistivity $\rho$ of Sr$_{1-x}$La$_x$RuO$_3$ for $x$ $\leq$ 0.5. The electrical resistivity of SrRuO$_3$ exhibits a metallic behavior over the entire temperature range  and shows a kink at the FM transition temperature of 161 K.  These features are in  agreement with previous measurements performed on  polycrystalline samples.\cite{shikano, neumeier} 
The kink anomally at $T^{*}$ in  $\rho$ becomes very small for $x=0.1$ (the inset of Fig. 5(a)) and then disappears upon further doping La.
This is consistent with the features of the specific heat at $\sim$$T^{*}$. With increasing the La concentration,  $\rho$ increases over the whole temperature range  for $x$ $\geq$ 0.2, and pronounced upturns develop at low temperature for $x$ $\geq$ 0.3. At the same time, $\rho$ at 5 K increases with increasing $x$ (Fig. 5(b)). Although the electrical resistivity is enhanced for $x$ $\geq$ 0.3, the relatively small magnitude of the $\rho$ values, even at low temperatures, suggests that this system preserves metallic characteristics up to $x$ = 0.5. In fact, the end material LaRuO$_3$ exhibits a metallic behavior.\cite{bouchard,sugiyama} In addition, the sensitive change in the magnitude of  $\rho$ at low temperatures is suggestive of the dominant contribution of the Ru 4$d$ electrons to the metallic conduction.  The increases in $\rho$ at low temperatures may be due to the grain boundary scattering which is often seen in the polycrystalline sintered samples. Another possible origin of the increases in $\rho$ is the random distribution of cations, which have different valencies and/or electronic correlation effects.  Such localization effects are expected to increase the electrical resistivity at low temperatures.\cite{lee}

A further indication of the itinerant character of the Ru 4$d$ states can be found in the $x$ variation of the effective moment for $x$ $\leq$ 0.5 estimated from the Curie-Weiss fit of the inverse susceptibilities, which is plotted in Fig. 5(c). Since Sr$^{2+}$ is substituted by La$^{3+}$, the same fraction of  Ru$^{4+}$ is expected to be substituted by Ru$^{3+}$ through electron transfer under the assumption that electrons are  localized at the ions. The Ru$^{4+}$ and Ru$^{3+}$ ions have effective moments of 2.83 and 1.73 $\mu_\mathrm{B}$, respectively,\cite{bouchard} and the calculated effective moment as a function of $x$ in this assumption is shown as a broken line in Fig. 5(c). We found that the effective moment obtained from the experiments does not show a monotonous decrease with $x$, which thus cannot be explained in terms of the simple ionic configurations of Ru$^{4+}$ and Ru$^{3+}$. In the itinerant electron model, by contrast, the Curie constant represents the stiffness of the longitudinal spin fluctuation and thus may be different from the prediction from the localized electron model.\cite{kawanaka}

\section{Discussion}

We observed relatively large $\gamma$ values over the entire $x$ range presently investigated. This indicates that the density of states around Fermi level is dominated by the Ru 4$d$ states.  In addition, the electrical resistivity also shows a metallic nature, and the $x$ dependence of the effective moment cannot simply be understood in terms of  the simple localized electron model. Thus, we naturally expect that the itinerant aspect of the Ru 4$d$ electrons should be taken into account in order to understand the magnetism of Sr$_{1-x}$La$_x$RuO$_3$. In general, the second-order FM transition ends at a tricritical point and changes into a first order transition by controlling external parameters, such as pressure for clean itinerant ferromagnets.\cite{belitz}  This type of phase diagram has been reported for several itinerant ferromagnets, such as ZnZn$_2$ and UGe$_2$, where the FM transition temperature is monotonously suppressed with increasing pressure and then shows a rapid drop after passing the tricritical point.\cite{taufour,kabeya} On the other hand, the $x$ variation of $T^{*}$  and the emergence of the cluster-glass state in Sr$_{1-x}$La$_x$RuO$_3$ are clearly incompatible with the above features, and these differences further indicate that the disorder caused by doping plays an important role in the magnetic properties in Sr$_{1-x}$La$_x$RuO$_3$. In addition, the significant frequency dependence of $T^{*}$ for $x$ $\geq$ 0.3 means that $T^{*}$ is no longer a real phase transition but rather a gradual freezing. The quantum critical point of the FM order is thus considered to be absent or smeared by the substitutions in Sr$_{1-x}$La$_x$RuO$_3$. This is consistent with the fact that a non-Fermi-liquid-like divergence is not observed in  $\chi'_\mathrm{ac}$, and the electronic contribution in the specific heat shows a Fermi-liquid temperature dependence at low temperatures over the entire $x$ range investigated in this study. Here, we wish to stress that the itinerant character and the spatially inhomogeneous magnetic order states revealed in the present system are generally conflicting properties, and the coexistence can be understood neither by a simple itinerant nor by a simple localized picture of electrons. In this sense, the Ru 4$d$ electrons in this system have a duality of itinerant and localized natures.   Recent investigations for dc-magetization of Sr$_{1-x}$Ca$_x$RuO$_3$ have also indicated this feature,\cite{demko} implying a common mechanism of the FM suppression underlying in Sr-site substituted SrRuO$_3$. In addition, the photoemission and optical conductivity measurements have recently revealed that the  duality of itinerant and localized natures appears even in pure SrRuO$_3$.\cite{shai,jeong}  

The  $\delta$ values estimated from the frequency dependence of $T^{*}$ indicate that the low temperature ordered phase for $x$ $\geq$ 0.3 can not be realized to a spin-glass but a cluster-glass state.  Furthermore, the present dc-magnetization curves measured in the field cooled conditions are clearly incompatible with those expected from the spin-glass system (Fig. 1(a)).  It is generally known that dc-magnetization in the field cooled condition of canonical spin-glass systems is nearly independent of temperature below $T^{*}$.\cite{mydosh} In contrast, the continuous enhancement of the dc-magnetization is frequently observed below $T^{*}$ in the cluster-glass system,\cite{mukherjee,marcano,marcano2,szlawska} and it closely resembles the present observations for $x$ $\geq$ 0.3. In particular, the overall temperature dependence of the dc-magnetization, frequency dependences of $T^{*}$ and amplitude  $\chi'_\mathrm{ac}$ at  $T^{*}$ are very similar to those of the CeNi$_{1-x}$Cu$_x$ system.\cite{marcano,marcano2}
 Neutron diffraction experiments revealed that CeNi$_{1-x}$Cu$_x$ exhibits a long range FM order at low temperatures for 0.3 $\leq$ $x$ $\leq$ 0.6. The ac-susceptibility shows a peak at much higher temperature than the onset of the long range FM order, and its frequency dependence suggests a cluster-glass order. No indication of the long-range FM ordering was detected by macroscopic measurements, such as ac-susceptibility and specific heat.  In order to understand this puzzling magnetic properties, the cluster-percolative scenario is proposed, in which a crossover from the cluster-glass to FM ordered states occurs. However, CeNi$_{1-x}$Cu$_x$ has a more localized character of the electrons than Sr$_{1-x}$La$_x$RuO$_3$, since its magnetism originates in localized 4$f$ electrons, and the relatively small Kondo temperature is realized in these alloys.\cite{marcano3} 

Finally, we discuss the $x$ dependence of  $E_\mathrm{a}$ and $T_0$ characterizing the cluster-glass properties. Both $E_\mathrm{a}$ and $T_0$ are usually expected to be reduced with increasing $x$ because of the suppression of the FM correlation by doping La, which likely leads to shrinkage of the magnetic cluster size as well as an increase of the inter-cluster distance. In fact, $T_0$ shows a decrease with  doping La in accord with our expectation. However, interestingly, $E_\mathrm{a}$ is nearly constant with increasing $x$, indicative of the cluster size being unchanged  by $x$. Recently, the nature of the cluster size distribution is argued on the basis of the optimal fluctuation theory combined with the finite size scaling technique.\cite{hrahsheh}
Let us focus on a small region in a sample where the total number of the Sr and La sites is $N$, and $L_\mathrm{RR}$ is its linear size. For a given $x$ value, the probability that the small region contains the  $N_\mathrm{La}$ number of La atoms is given by the binomial distribution $P(N_\mathrm{La})=\binom{N}{N_\mathrm{La}}x^{N_\mathrm{La}}(1-x)^{N-N_\mathrm{La}}$. Here, we assume that the small region becomes a magnetically ordered state when its local La concentration $x_\mathrm{loc}=N_\mathrm{La}/N$ is smaller than the threshold value $x_\mathrm{c}(L_\mathrm{RR})$. It was actually argued that such a local order can be stabilized by the overdamped cluster dynamics.\cite{vojta,leggett} According to the finite size scaling, the functional form of  $x_\mathrm{c}(L_\mathrm{RR})$ is given by $x_\mathrm{c}(L_\mathrm{RR})=x_\mathrm{c}^0 - DL_\mathrm{RR}^{-\phi}$, where $x_\mathrm{c}^0$ is the critical concentration for the bulk system, $D$ is a constant, and $\phi$ is the finite-size shift exponent. This equation implies the presence of a lower limit of the cluster size given by $L_\mathrm{min}=(D/x_\mathrm{c}^0)^{-\phi}$. The magnetization value can be estimated from simple integration of  all of the regions that show local magnetic order, and the magnetization is dominated by the minimum size clusters  for the large $x$ region.\cite{hrahsheh} In this senario, therefore, the $x$ independent $E_\mathrm{a}$ values observed for $x$ $\geq$ 0.3 could be related to the activation energy of the minimum size clusters.

Though above discussions have been made, further experimental studies using microscopic probes, such as $\mu$SR and neutron scattering experiments, are necessary to have a comprehensive understanding of the magnetic properties of Sr$_{1-x}$La$_x$RuO$_3$, and these microscopic measurements are now in progress.

\section{Conclusions}
We have performed  dc-magnetization, ac-susceptibility, specific heat, and electrical resistivity measurements on Sr$_{1-x}$La$_x$RuO$_3$  in order to investigate its magnetic and electronic properties.  We found that the FM ordered state is strongly suppressed by  La substitution. Furthermore, the onset of FM order, determined by the peak in  $\chi'_\mathrm{ac}$, shows a significant frequency dependence for $x$ $\geq$ 0.3. The estimated initial frequency shifts $\delta$ are comparable to those of cluster-glass systems. The frequency dependence of  $T^{*}$ is well reproduced by the Vogel-Fulcher law, and precise analyses revealed that the activation energy $E_\mathrm{a}$ of the FM clusters does not show a significant change by $x$, while the Vogel-Fulcher temperature $T_0$ is reduced. 
The former indicates that the cluster size is nearly independent of $x$.

The linear specific heat coefficient $\gamma$ increases accompanying the suppression of magnetic order. The relatively large $\gamma$ values are ascribed to a presence of the Ru 4$d$ states at  Fermi level. Although the electrical resistivity shows pronounced increases at low temperatures for $x$ $\geq$ 0.3, small electrical resistivity values, even at low temperatures, reflect a metallic nature of the electrons in these compounds. We also found that the $x$ variation of the effective moment estimated from the Curie-Weiss fit cannot be understood in terms of a simple configuration of Ru ions where the 4$d$ electrons are localized at Ru ions. These results suggest that the magnetism of Sr$_{1-x}$La$_x$RuO$_3$ is attributed to the itinerant characteristics of the Ru 4$d$ electrons. The coexistence of the itinerant character and the significant spatial inhomogeneity yielding the cluster-glass behavior can be understood neither by a simple itinerant picture nor by a simple localized picture of the Ru 4$d$ electrons.  It is challenging to resolve this puzzling issue, and thus, further experimental and theoretical studies are needed.

\begin{acknowledgment}
We would like to thank Y. Nishihara for helpful discussions. This work was partially supported by Grants-in-Aid for Scientific Research on Innovative Areas ``Heavy Electrons" and ``Topological Quantum Phenomena" from the Ministry of Education, Culture, Sports, Science and Technology of Japan. 
\end{acknowledgment}

\end{document}